\documentclass[a4paper,11pt]{article}
\pdfoutput=1 

\usepackage{jinstpub} 

\usepackage{float}
\floatstyle{plaintop}
\restylefloat{table}

\usepackage{caption} 
\usepackage{hyperref}

\usepackage[parfill]{parskip} 

\usepackage{cleveref} 
\usepackage[utf8]{inputenc} 
\usepackage{isotope} 
\usepackage{lineno} 
\usepackage{tablefootnote} 
\usepackage{textcomp} 
\usepackage[normalem]{ulem} 
\usepackage[usenames,dvipsnames]{xcolor} 
\usepackage{xfrac} 
\usepackage{xspace} 

\newboolean{showedits}
\setboolean{showedits}{true} 
\newboolean{showcomments}
\setboolean{showcomments}{true} 

\newcommand{\tubii}{TUB\lowercase\expandafter{\romannumeral2}\xspace} 

\ifthenelse{\boolean{showedits}}
{
	\newcommand{\del}[1]{\textcolor{red}{\sout{#1}}}   
}{
	\newcommand{\del}[1]{}   
}

\newcommand{\id}[1]{$-$ \textsc{\rversion} $-$}

\ifthenelse{\boolean{showcomments}}
{
	\newcommand{\ob}[2]{ 
		{\noindent\colorbox{Orange}
			{\bfseries\sffamily\scriptsize\textcolor{white}{#1}}}
		{\textcolor{RedOrange}
			{\sf\small$\blacktriangleright${#2}$\blacktriangleleft$}}
	}
	\newcommand{\bb}[2]{ 
		{\noindent\colorbox{MidnightBlue}
			{\bfseries\sffamily\scriptsize\textcolor{white}{#1}}}
		{\textcolor{MidnightBlue}
			{\sf\small$\blacktriangleright${#2}$\blacktriangleleft$}}
	}
	\newcommand{\rb}[2]{
		{\noindent\colorbox{BrickRed}
			{\bfseries\sffamily\scriptsize\textcolor{white}{#1}}}
		{\textcolor{BrickRed}
			{\sf\small$\blacktriangleright${#2}$\blacktriangleleft$}}
	}
	\newcommand{\gb}[2]{ 
		{\noindent\colorbox{Green}
			{\bfseries\sffamily\scriptsize\textcolor{white}{#1}}}
		{\textcolor{Green}
			{\sf\small$\blacktriangleright${#2}$\blacktriangleleft$}}
	}
	
}
{
	\newcommand{\ob}[2]{}
	\newcommand{\rb}[2]{}
	\newcommand{\bb}[2]{}
	\newcommand{\gb}[2]{}
	
}

\ExplSyntaxOn
\newcommand{\latinabbrev}[1]{
  \peek_meaning:NTF .
  {#1\xspace}
  {
  	#1.\xspace
  }
}

\newcommand{\latinabbrevstyled}[1]{
  \peek_meaning:NTF .
  {\emph{#1}\xspace}
  {
  	\emph{#1.}\xspace
  }
}
\ExplSyntaxOff

\newtoggle{people} 
\togglefalse{people}

\title{\boldmath Development, characterisation, and deployment of the SNO+ liquid scintillator}
\collaboration{
    \includegraphics[height=17mm]{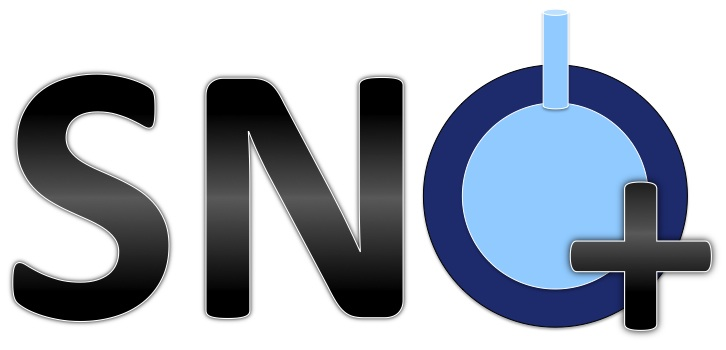}\\
    [6pt] SNO+ collaboration 
}

\author[a]{\bf M.\,R.\,Anderson}
\author[b]{\bf S.\,Andringa}
\author[c]{\bf L.\,Anselmo}
\author[d]{\bf E.\,Arushanova}
\author[a]{\bf S.\,Asahi}
\author[e,f,g]{\bf M.\,Askins}
\author[h]{\bf D.\,J.\,Auty}

\author[d,i]{\bf A.\,R.\,Back}
\author[j]{\bf Z.\,Barnard}
\author[b,k,l,m]{\bf N.\,Barros}
% ...

\affiliation[a]{\it Queen's University, Department of Physics, Engineering Physics \& Astronomy, Kingston, ON K7L 3N6, Canada}
\affiliation[b]{\it Laborat\'{o}rio de Instrumenta\c{c}\~{a}o e  F\'{\i}sica Experimental de Part\'{\i}culas (b), Av. Prof. Gama Pinto, 2, 1649-003, Lisboa, Portugal}
\affiliation[c]{\it SNOLAB, Creighton Mine \#9, 1039 Regional Road 24, Sudbury, ON P3Y 1N2, Canada}
\affiliation[d]{\it Queen Mary, University of London, School of Physics and Astronomy,  327 Mile End Road, London, E1 4NS, UK}
\affiliation[e]{\it University of California, Berkeley, Department of Physics, CA 94720, Berkeley, USA}
\affiliation[f]{\it Lawrence Berkeley National Laboratory, 1 Cyclotron Road, Berkeley, CA 94720-8153, USA}
\affiliation[g]{\it University of California, Davis, 1 Shields Avenue, Davis, CA 95616, USA}
\affiliation[h]{\it University of Alberta, Department of Physics, 4-181 CCIS,  Edmonton, AB T6G 2E1, Canada}
\affiliation[i]{\it University of Sussex, Physics \& Astronomy, Pevensey II, Falmer, Brighton, BN1 9QH, UK}
\affiliation[j]{\it Laurentian University, Department of Physics, 935 Ramsey Lake Road, Sudbury, ON P3E 2C6, Canada}
\affiliation[k]{\it Universidade de Lisboa, Faculdade de Ci\^{e}ncias (FCUL), Departamento de F\'{\i}sica, Campo Grande, Edif\'{\i}cio C8, 1749-016 Lisboa, Portugal}
\affiliation[l]{\it University of Pennsylvania, Department of Physics \& Astronomy, 209 South 33rd Street, Philadelphia, PA 19104-6396, USA}
\affiliation[m]{\it Technische Universit\"{a}t Dresden, Institut f\"{u}r Kern und Teilchenphysik, Zellescher Weg 19, Dresden, 01069, Germany}
% ...

\emailAdd{benjamin.tam@queensu.ca}
\emailAdd{ljpickard@ucdavis.edu}

\abstract{
A liquid scintillator consisting of linear alkylbenzene as the solvent and 2,5-diphenyloxazole as the fluor was developed for the SNO+ experiment. 
This mixture was chosen as it is compatible with acrylic and has a competitive light yield to pre-existing liquid scintillators while conferring other advantages including longer attenuation lengths, superior safety characteristics, chemical simplicity, ease of handling, and logistical availability.
Its properties have been extensively characterized and are presented here.
This liquid scintillator is now used in several neutrino physics experiments in addition to SNO+.
}


\arxivnumber{2011.12924} 

\begin{document}

    \pagenumbering{Alph}
    \maketitle
    \flushbottom 

	\graphicspath{{./}{figures/introduction/}}
\section{Introduction}
\label{sec:introduction}
\pagestyle{myplain}\pagenumbering{arabic}
\setcounter{page}{1}

Scintillation counters continue to be among the most common particle detectors. Scintillators are materials that emit light following excitation by ionising radiation \cite{Birks1964}; in particle detectors, this light is typically observed with photomultiplier tubes (PMTs).

Organic scintillators --- the favoured scintillation medium in large liquid particle detectors --- are aromatic solvents with an emission mechanism that depends on the excitation and subsequent de-excitation of benzene rings in various molecules \cite{Brooks1959}. Liquid scintillators are typically ``cocktails" that include fluors and wavelength shifters in order to maximize light yield and transparency, optimize emission times, minimize self-absorption, and tune the emission spectra to match the quantum efficiency curves of the observing PMTs \cite{Brooks1979}. Recent neutrino experiments that have successfully deployed large volumes of organic liquid scintillator include Borexino, which used 1,2,4-trimethylbenzene (i.e. pseudocumene, PC) \cite{Elisei1997} and KamLAND, which used PC diluted in dodecane \cite{Yoshida2010}. Both Borexino and KamLAND used 2,5-diphenyhloxazole (PPO) as the fluor.

\subsection{The SNO+ Experiment}
\label{sec:snoplus}

SNO+ is a large-scale, multipurpose neutrino physics experiment. The detector is situated 2\,km underground at the SNOLAB facility located in the Creighton mine near Sudbury, Canada. The basic infrastructure of the SNO+ detector was inherited from the SNO experiment, including the 6-m radius spherical acrylic vessel (AV) that serves to contain the active target \cite{Andringa2016}. A detailed description of the SNO+ detector is given in \cite{detectorpaper}.

Acrylic is not compatible with existing widely-used liquid scintillators such as PC. This incompatibility motivated the SNO+ collaboration to search for a new option that would not only be compatible with acrylic, but have a competitive light yield while being safer to handle. This search culminated in the identification of a cocktail consisting of linear alkylbenzene (LAB) as the solvent and PPO as the fluor. Since its development and adoption by the SNO+ collaboration, LAB-based liquid scintillators have been successfully deployed in a number of neutrino detectors such as Daya Bay \cite{DYB} and RENO \cite{Park2010}, and will be used in future neutrino experiments including JUNO \cite{Cao2019}. LAB-based scintillators are also used in veto detectors for current dark matter experiments such as COSINE-100 \cite{cosine100}, and are considered for use in future dark matter experiments such as SABRE \cite{sabre}.

After an extensive physics campaign operating as an ultrapure water Cherenkov detector \cite{Anderson2019, Anderson2019_2, Anderson2020}, liquid scintillator has now been deployed in the SNO+ detector. The physics program in this phase of the experiment requires precise characterization of the scintillator to facilitate modelling of the detector response. The physics program also required the scintillator to be of ultra high purity in order to limit radiological backgrounds and minimize light attenuation \cite{detectorpaper}; a purification plant was built to purify the liquid scintillator prior to deployment into the SNO+ detector.

This paper discusses the SNO+ liquid scintillator: its conception and development, the characterization of its properties, and the purification and deployment techniques used.
	\begingroup	
	\let\clearpage\relax
	\section{Scintillator Development}
\label{sec:development}

The incompatibility of acrylic with commonly used scintillating solvents motivated the SNO+ collaboration to search for a new cocktail. Diluted and mixed solvents, such as the 80\% dodecane + 20\% PC + PPO (1.52\,g/L) mixture used in the KamLAND experiment, have been demonstrated to be sufficiently compatible with acrylic \cite{Yoshida2010}. However, there was a desire to avoid using PC in SNO+ due to safety concerns --- PC is highly volatile, toxic, and has a low flash point. This prompted the development of a new liquid scintillator with a light yield competitive with PC, while prioritising acrylic compatibility and superior safety handling considerations. As the AV is submerged in a water-filled cavity that acts as external shielding, a liquid scintillator density close to 1\,g/cm$^{3}$ was also desirable.

\subsection{Solvent}

There were seven solvents identified as potential candidates for the SNO+ liquid scintillator. These are listed in Table \ref{tab:solvents}.

\begin{table}[h!]
\centering
\begin{tabular}{ ||c|c|c|| }
\hline
 Solvent & Density (g cm$^{-3}$) & Flash point ($^{\circ}$C)\\
\hline
 Pseudocumene (PC) & 0.889 & 44 \\ 
 Phenylcyclohexane (PCH) & 0.950 & 98\\  
 Linear alkylbenzene (LAB) & 0.856 & 143\\
 Di-isopropylnaphthalene (DIN) & 0.960 & 140 \\
 Phenyl-o-xylyletane (PXE) & 0.985 & 167 \\
 1-Methylnaphthalene (1-MN) & 1.020 & 82 \\
 Dodecylbenzene (DCB) & 0.870 & 140 \\
 \hline
\end{tabular}
\caption{Solvent candidates investigated for the SNO+ liquid scintillator.}
\label{tab:solvents}
\end{table}

Acrylic compatibility, the chief factor in the initial evaluation of the various solvents, was determined by submerging 80 cast acrylic cubes ($6.35$\,mm$\,\times\,6.35$\,mm$\,\times\,6.35$\,mm) in 175\,mL of each candidate solvent prior to stirring with a magnetic teflon rod. A Photon Technology International (PTI) QuantaMaster\textsuperscript{TM} fluorescence spectrometer subsequently measured Mie scattering caused by any acrylic particulates suspended in each of the acrylic-exposed solvents; such particulates are indicative of chemical degradation of the cubes. Measurements were taken 1--2 days apart for a period of $\sim$2\,weeks. Poor compatibility excluded DCB and PCH as candidates. As expected, PC was also not compatible with acrylic. The remaining solvents exhibited varying --- from none to tolerable --- degrees of degradation, and were used in further selection testing.

Light yield was another major factor during selection. To determine the light yields of the remaining solvents, each were doped with PPO (2\,g/L) and excited with a 10\,mCi $^{125}$I source. The resulting emission spectra were measured using a PTI QuantaMaster\textsuperscript{TM} fluorescence spectrometer; the intensities between 350--425\,nm were integrated and normalized to measurements using PC to give a relative pulse height (RPH). 1-MN was measured to have a poor light yield, though it is known that the light yield of 1-MN is highly dependent on purity. Furthermore, 1-MN was previously found to need careful treatment due to issues of robustness \cite{Goldstein1964}. Since 1-MN exhibited no obvious advantages, these factors excluded this solvent from further selection testing.

As other experiments have deployed solvents diluted in dodecane to improve acrylic compatibility \cite{Aberle2011}, this test was repeated for each of the remaining candidates after dilution with dodecane at various concentrations. The measurements were compared to a sample of KamLAND scintillator, as shown in Figure \ref{fig:dilution_ly}. 

\begin{figure}[htp]
\centering
	\includegraphics[width=0.8\textwidth,trim={0 0 0 0},clip]{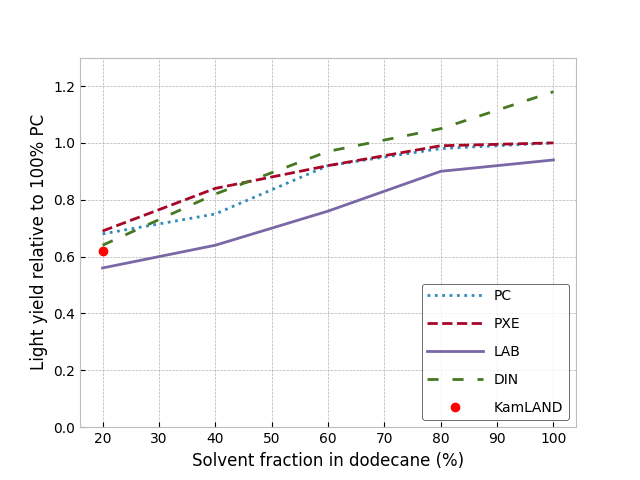}
	\caption{The effect of dodecane dilution on the light yield of the remaining solvent candidates. The red marker indicates a measurement made of the KamLAND scintillator, which has a 20\% solvent fraction in dodecane.}
	\label{fig:dilution_ly}
\end{figure}

DIN appeared to be a viable choice due to its competitive acrylic compatibility, high intrinsic light yield, and comparable density to water. However, compared to LAB, DIN has a shorter attenuation length of $3.3$--$4.4$\,m at 430\,nm \cite{song2013} which is comparable to the scale of the SNO+ detector; this would result in a significantly lower overall light collection.\footnote{Preliminary measurements of LAB demonstrated mean attenuation lengths of over 20~m. More extensive measurements were later performed on LAB, and are discussed in Sections \ref{sec:scattering} and \ref{sec:emission}.}  Mixtures which minimized acrylic damage while maintaining competitive light yields were also considered --- particularly those involving PXE. However, PXE is incompatible with acrylic when undiluted and has been demonstrated to soften acrylic above a solvent fraction of 20\% in dodecane \cite{aberle2011_b}. Compared to LAB, no combination of solvents and/or dilutions conferred sufficient advantages to justify the increased cocktail complexity.

LAB was determined to be the superior compromise between acrylic compatibility, light yield, cocktail simplicity, and safety. Furthermore, LAB is a single-component solvent, has a near-identical index of refraction to acrylic, and is easy to handle due to its non-toxicity and mild odour. Commercial LAB has minimal Rayleigh scattering and high transparency, with attenuation lengths at relevant wavelengths of over 10\,m even prior to purification. LAB is also a logistically favorable choice due to the relative proximity of SNOLAB to the CEPSA Qu\'imica B\'ecancour (formerly known as Petresa Canada) linear alkylbenzene facility in B\'ecancour, Quebec.

Once LAB was identified as the premier solvent, an extensive investigation was initiated to confirm its acrylic compatibility, following ASTM D543 ``Standard Practices for Evaluating the Resistance of Plastics to Chemical Reagents''. After 100 months of testing at elevated stresses and strains, it was found that LAB has a negligible chemical effect on the acrylic used in the SNO+ experiment. Details of the long-term acrylic compatibility tests and the experimental design are given in \cite{Wright2009} with the most recent results given in \cite{Bartlett2018}.

\subsection{Fluor}

PPO has been the dominant choice of fluor in liquid scintillators for decades \cite{Wunderly1990}. Two additional solutes, 2-(4-Biphenyl)-5-phenyloxazole (BPO) and 2-(1-Naphthyl)-5-phenyloxazole (NPO), were also identified as potential options. Furthermore, phenyls such as p-terphenyl (another common fluor), along with p-quinquephenyl and p-sexiphenyl (both less common), were briefly considered but excluded due to their low solubility in LAB. Although BPO and NPO have slightly higher intrinsic light yields, PPO was ultimately chosen as the fluor due to its low cost, wide availability, and the extensive existing literature as a result of its widespread use.

\subsection{Future development}

In the initial scintillator development phases, metal-loading capabilities were an independent consideration and did not impact the selection process. A technique was ultimately established to load natural Te into the scintillator in order to allow for the study of neutrinoless double beta decay with $^{130}$Te. To mitigate light absorbed by Te, 1,4-bis(2-methylstyryl)benzene (bis-MSB) can be added to the scintillator as an additional secondary wavelength shifter. Bis-MSB is known to perform well with PPO as a fluor-shifter mixture. The Te loading process will be discussed in a forthcoming paper.

	\section{The SNO+ Liquid Scintillator}
\label{sec:cocktail}

The final SNO+ liquid scintillator (``the scintillator cocktail'') was determined to be 2\,g PPO per 1\,L LAB. Increasing the concentration of PPO in LAB increases both the primary light yield and self-absorption. The effect of PPO concentration on primary light yield was published in \cite{Cumming2018}. The effect of PPO concentration on self-absorption is detector-dependent and studied using dedicated Monte Carlo simulations. With the SNO+ detector, the increase in light collection with increasing amounts of PPO would start to plateau at concentrations higher than 2\,g/L.

LAB acts as both the solvent and primary absorber. The carbon chain attached to the phenyl group varies from 9--14 (>95\% 9--12) carbons in length, as seen in Figure \ref{fig:structures} (left); each chain length has subtle emission spectra differences.
As discussed in Section \ref{sec:emission}, the presence of a fluor mitigates self-absorption of LAB by transferring energy from LAB to PPO (primarily via non-radiative F\"{o}rster resonant energy transfer), resulting in light being emitted at longer wavelengths. The structural formula for PPO can be seen in Figure \ref{fig:structures} (right).

\begin{figure}[htp]
    \centering
    \includegraphics[width=0.47\textwidth]{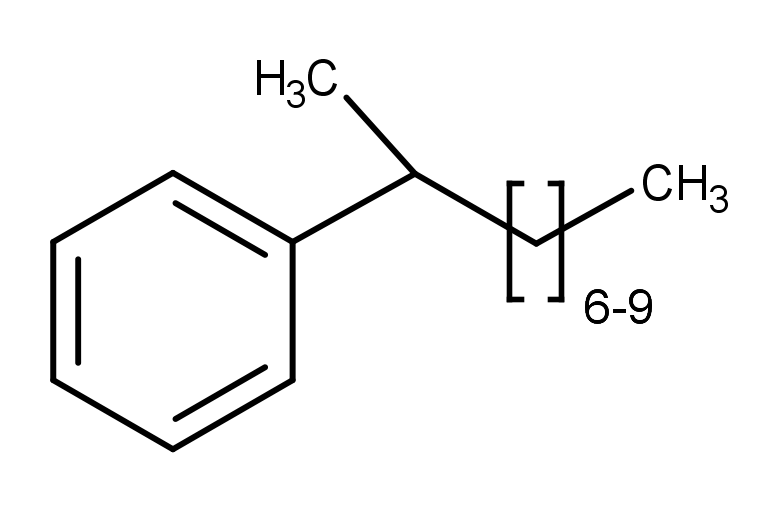}
	\includegraphics[width=0.47\textwidth]{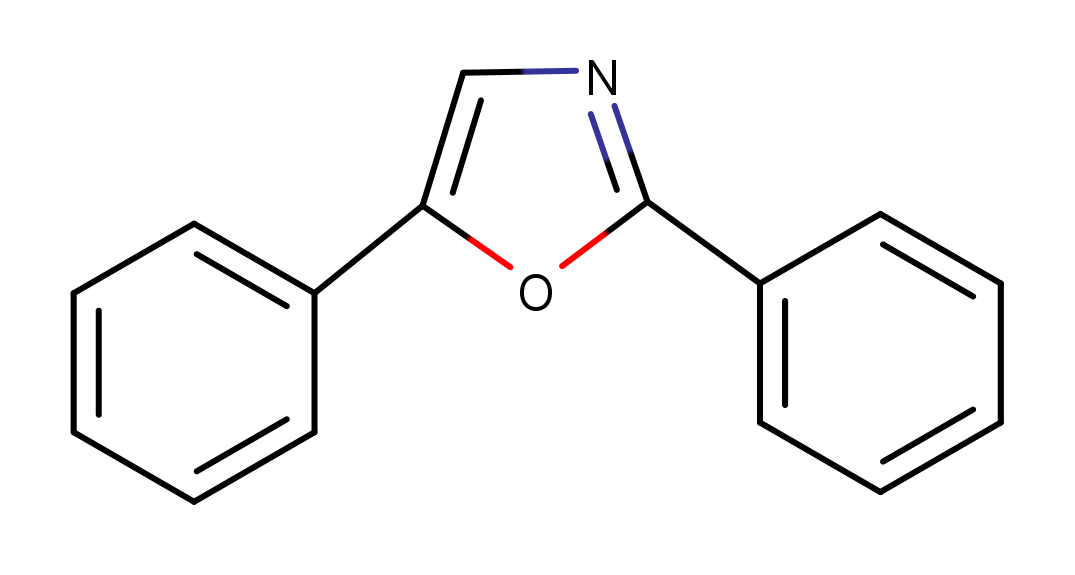}
    \caption{Structural formulae of the two primary components of the SNO+ scintillator cocktail. Left: LAB is a phenyl group attached to a carbon chain varying between 9--14 (>95\% 9--12) carbons in length. Right: PPO is the fluorophore of the liquid scintillator cocktail.}%
    \label{fig:structures}%
\end{figure}

The scintillator cocktail exhibits acrylic compatibility and competitive light yields with respect to existing liquid scintillators, while maintaining long attenuation lengths, improved safety characteristics, chemical simplicity, ease of handling, and logistical availability to the SNO+ site. 

As part of the final selection, a preliminary comparison of the decay time was performed between the unpurified scintillator cocktail, the scintillator used in the KamLAND experiment, and PC + \,2g/L PPO (the scintillator formula used in the Borexino experiment). These results are presented in Figure \ref{fig:comparison}. The scintillation times were deduced using a single exponential fit on the prompt component of the time spectrum after UV excitation. A more thorough analysis of the scintillator cocktail timing was subsequently undertaken and is presented in Section \ref{sec:timing}.

\begin{figure}[htp]
\centering
	\includegraphics[width=0.75\linewidth]{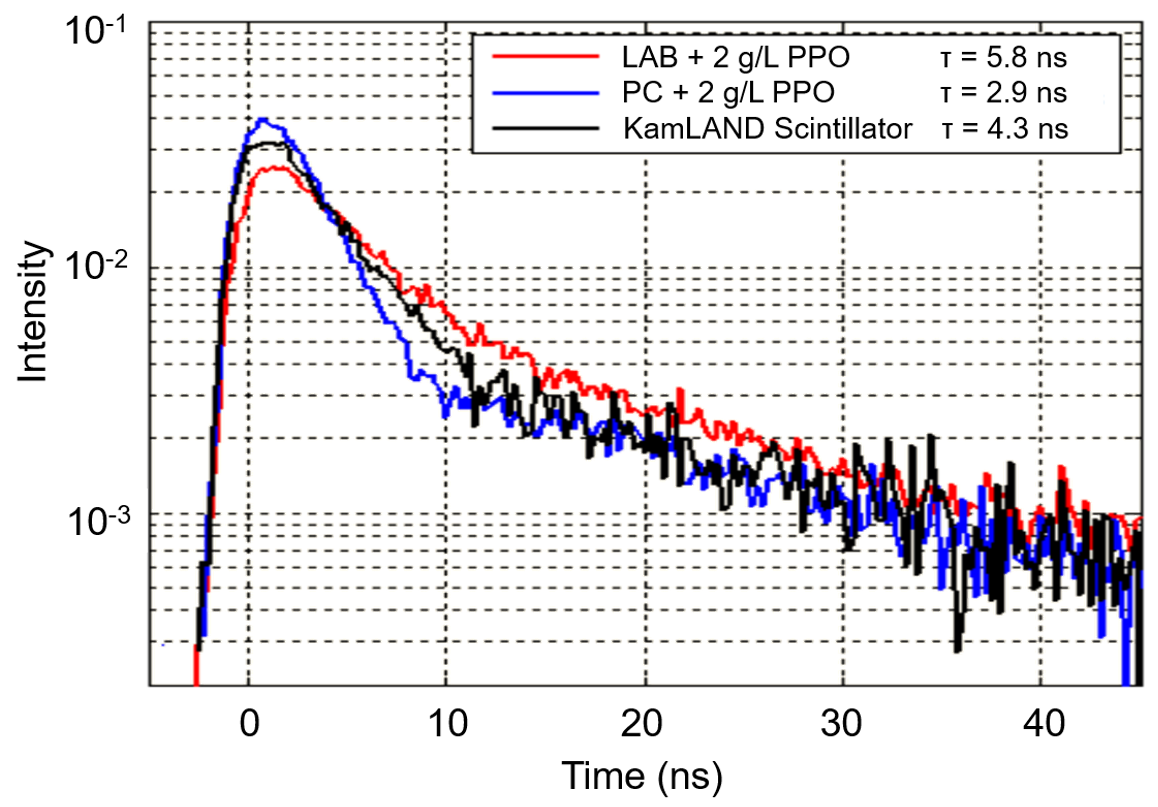}
	\caption{Comparison of scintillation decay times for the KamLAND scintillator, PC + 2\,g/L PPO, and LAB + 2\,g/L. The time constant $\tau$ for each was determined using a single exponential fit on the prompt component.}
	\label{fig:comparison}
\end{figure}

	\section{Scintillator Properties}
\label{sec:scintprop}

The properties of the scintillator cocktail have been extensively characterized and subsequently incorporated into the SNO+ Monte Carlo framework to ensure that the detector is accurately and precisely modelled \cite{Andringa2016}.

\subsection{Light Yield}
\label{sec:lightyield}

As with all scintillation detectors, characterization of the light yield was required to understand the energy resolution of the SNO+ experiment. Two different methods to determine the light yield were carried out: irradiation using an $^{125}$I source with the light output measured using a PMT-based spectrometer, and irradiation using a $^{90}$Sr source with the light output measured using a PMT.

\paragraph{$^{125}$I irradiation}
The light yield of the scintillator cocktail was quantified through comparison to PC + 2\,g/L PPO. Each were poured into a 1\,$\times$\,1\,cm quartz cuvette located in a dark box, and subsequently irradiated with 35\,keV X-rays from an $^{125}$I source. The emission spectra were measured using a PMT-based spectrometer. The relative light yield was determined by integrating their spectra between 350--425\,nm. After accounting for the $\sim$5\,cm attenuation lengths and the density differences between the solvents, LAB was found to have a relative light yield of $\sim$96\% compared to PC with equivalent PPO loading.

The light yield of PC + PPO\,(at 1.5\,g/L) was determined by the Borexino collaboration to be 11500\,$\pm$\,1000\,photons/MeV \cite{Elisei1997}. Based on measurements that were performed, the light yield needed to be scaled by (7.5$\pm$0.5)\% to obtain a 2.0\,g/L equivalent loading. Accounting for the $\sim$96\% relative measurement, the absolute light yield of the SNO+ scintillator cocktail was determined to be 11900\,$\pm$\,1100\,photons/MeV.

\paragraph{$^{90}$Sr irradiation}
Irradiation with a 0.1\,$\mu$Ci $^{90}$Sr source provided a complementary light yield measurement of the scintillator cocktail. The source was embedded in a $3.5$\,cm\,$\times$\,$3.5$\,cm\,$\times$\,$3.5$\,cm acrylic cube with a hollowed out, 2\,cm diameter, $\sim$11\,mL cylindrical sample cavity. The size of the cube was optimized to reduce self-absorption. $^{90}$Sr decays to $^{90}$Y, emitting a 0.546\,MeV $\beta^{-}$ that is fully absorbed by the acrylic. However, $^{90}$Y decays to $^{90}$Zr, emitting a 2.28\,MeV $\beta^{-}$ which penetrates through the acrylic and ionizes the sample. A 1" Hamamatsu R7600-200 HQE PMT was used to trigger an acquisition, while three 12" Hamamatsu R11780 HQE PMTs were used to collect the scintillation light. This setup was located in a 2\,m\,$\times$\,2\,m\,$\times$\,0.9\,m dark box surrounded by a Helmholtz coil and FINEMET\textsuperscript{\textregistered} shielding to reduce the effect of the Earth's magnetic field.

For each triggered event, the total charge measured by all three 12" PMTs was used to build a charge spectrum. 
The accumulated spectrum was compared to a simulation performed using a Geant4-based Monte Carlo framework. The experiment and simulation was also carried out for Cherenkov light by placing water in the sample container, which was used to normalize the integrated light collection. The light yield of the scintillator cocktail was determined to be 10830\,$\pm$\,570 photons/MeV. The measured and simulated spectra for both LAB and water are shown in Figure \ref{fig:lightyield}.
As oxygen is known to be responsible for quenching effects in scintillators \cite{OKeeffe2011}, this method was repeated after the sample was de-oxygenated through 30\,minutes of sparging with N$_{2}$. The light yield of the scintillator cocktail after de-oxygenation was increased to 11920\,$\pm$\,630\,photons/MeV. This is in good agreement with the $^{125}$I X-ray assay. The systematic uncertainties for this test are presented in Table \ref{table:ly_uncertainties}. 

\begin{figure}[htp]
  \centering
    \includegraphics[width=0.8\linewidth]{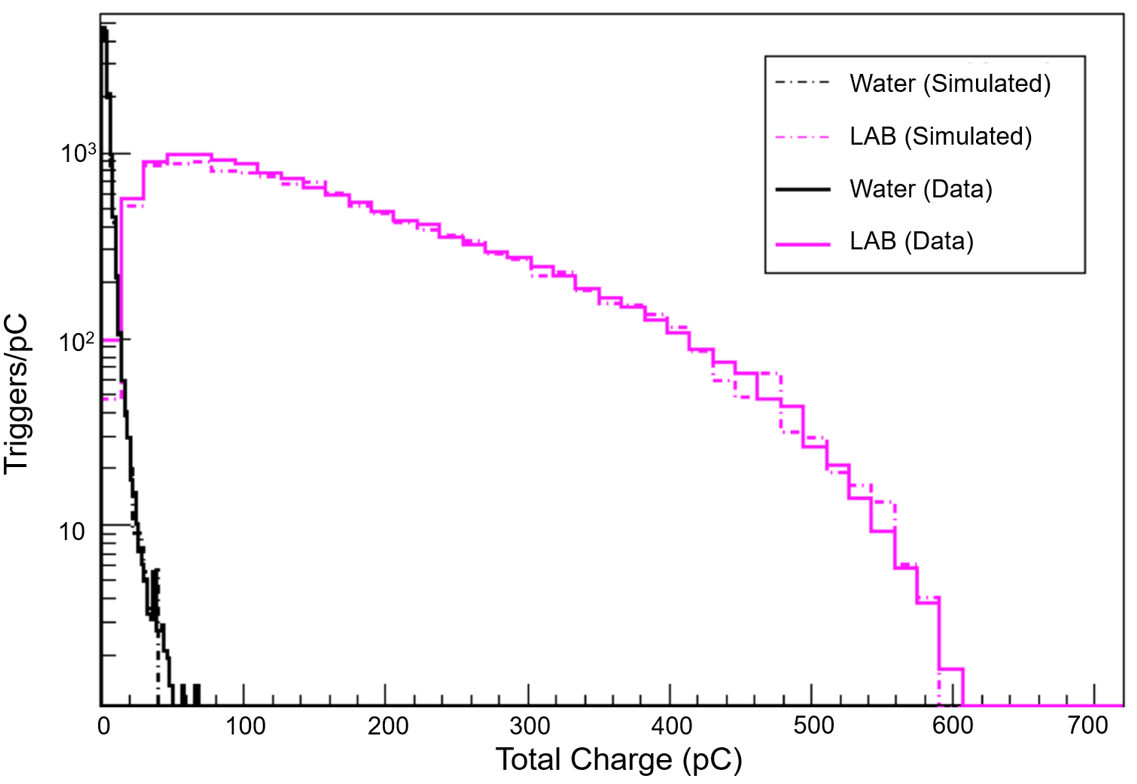}
  \caption{The measured charge spectra for the scintillator cocktail compared to a simulated best fit. The integrated light collection was normalized to Cherenkov light emitted by a water sample placed in the same sample container.}
      \label{fig:lightyield}
\end{figure}

\begin{table}[htp]
\centering
\begin{tabular}{||c|c||} 
 \hline
 Systematic Uncertainty & Value \\ 
  \hline
PMT distance to source & $\pm$0.5\,cm\\
Acrylic window thickness & $\pm$0.5\,mm\\
Source disc thickness & $\pm$0.25\,mm\\
Birks' constant & $\pm$4\%\\
PMT quantum efficiency & $\pm$3.5\%\\
\hline
Overall systematic uncertainty & $\pm$5.3\%\\
 \hline
\end{tabular}
\caption{A breakdown of systematic uncertainties for the $^{90}$Sr-irradiated light yield measurement.}
\label{table:ly_uncertainties}
\end{table}

\subsubsection{Quenching}

There are a number of non-radiative de-excitations or transitions that do not result in light generation. Such quenching effects reduce the light yield of the scintillator and are modeled through the Birks' equation:
\begin{equation}
    \frac{dL}{dx} = S \frac{\frac{dE}{dx}}{1+k_B\frac{dE}{dx}}
\end{equation}
where $\frac{dL}{dx}$ is the fluorescent energy emitted per unit path length, $\frac{dE}{dx}$ is the specific energy loss of the charged particle, and S is the scintillation efficiency which provides a normalization. $k_B$ is a material-specific Birks' constant.

Various quenching effects and the Birks' constant of the scintillator cocktail has been extensively studied by the SNO+ collaboration. Recent measurements of the Birks' constant are in the range of 66\,$\pm$\,16 to 76\,$\pm$\,3\,$\mu$m/MeV for $\alpha$-particles~\cite{krosigk2016}, 94\,$\pm$\,2 to 98\,$\pm$\,3\,$\mu$m/MeV for protons~\cite{krosigk2013}, and $\sim$74\,$\mu$m/MeV for electrons~\cite{tseung2011}. Additionally, the quenching of the scintillator cocktail has a temperature dependence that varies based on the particle type \cite{sorensen2018}. These measurements have been corroborated through multiple other studies \cite{krosigk2015,sorensen2016}.

\subsection{Scattering}
\label{sec:scattering}
The attenuation of photons via scattering was measured in order to accurately model SNO+ in simulations. Of the three main scattering mechanisms --- Rayleigh, Mie, and Raman --- Rayleigh scattering (in which the photon wavelength is much larger than the scattering molecule) dominates due to the size of the LAB and PPO molecules.\footnote{Rayleigh scattering is the dominant mechanism only if the scintillator is free of impurities with large molecular sizes, as is the case for the scintillator cocktail after purification.} The Rayleigh ratio and isotropic nature of the scattered light was assessed to understand this effect \cite{Satoko2012}. 

The Rayleigh ratio is the relationship between the intensities of incident and scattered light, and can be used to determine mean attenuation lengths. The Rayleigh ratio for LAB was determined by comparing its scattered light intensity with those of ultrapure water, acetone, cyclohexane, and toluene. This measurement was performed using a PTI QuantaMaster\textsuperscript{TM} fluorescence spectrometer after the solvents were placed in a 1\,$\times$\,1\,cm quartz cuvette within a dark box and illuminated with a Xe lamp monochromatized to 546\,nm. The Rayleigh ratio of LAB corresponding to each solvent was determined using
\begin{equation}
R_{LAB} = R_s \frac{I_{LAB}}{I_s} \frac{n_{LAB}}{n_s} A,
\label{eq:Rayleigh}
\end{equation}
where $R_{LAB}$ and $R_s$ are the Rayleigh ratio of the LAB and solvent, $I_{LAB}$ and $I_s$ are the measured intensities of the LAB and solvent, and $n_{LAB}$ and $n_s$ are the refractive index of LAB and solvent.\footnote{The refractive indices of the solvents were taken from \cite{Wahid1995}, while the refractive index for LAB was provided by the manufacturer, CEPSA Qu\'imica B\'ecancour Inc.} $A$ is a correction factor implemented to account for the differences in refractive indices between the solvent and the quartz cuvette. 

As seen in Table \ref{table:rayleigh_ratio}, the intensity of light scattered off of LAB was compared to that of each solvent to determine the Rayleigh ratio, with an average of (16.60\,$\pm$\,3.14)\,$\times$\,10$^{-6}$\,cm$^{-1}$ between the four solvents. Calculated from the Rayleigh scattering cross section, the resulting mean scattering length was determined to be 71.90\,$\pm$\,13.6\,m at 546\,nm.

\begin{table}[htp]
\centering
\begin{tabular}{||c| c| c| c|| } 
 \hline
 Solvent & Rayleigh ratio & Correction Factor & Corresponding Rayleigh ratio of LAB \\
  & (cm$^{-1}$) \cite{Wahid1995} & & (cm$^{-1}$) \\
  \hline
  Ultrapure Water & 1.05 & 21 & (16.63\,$\pm$\,2.99)\,$\times$\,10$^{-6}$\\
  Acetone & 4.47 & 5 & (16.79\,$\pm$\,3.14)\,$\times$\,10$^{-6}$\\
  Cyclohexane & 4.68 & 4 & (16.46\,$\pm$\,3.00)\,$\times$\,10$^{-6}$\\
  Toluene & 20.5 & 1 & (16.61\,$\pm$\,3.45)\,$\times$\,10$^{-6}$\\
 \hline
\end{tabular}
\caption{Using the Rayleigh ratio of known solvents, the Rayleigh ratio of LAB was determined by comparing the intensity of scattered light between LAB and each solvent.  The correction factor accounts for discrepancies in refractive indices between the solvent and quartz cuvette used for the measurements.}
\label{table:rayleigh_ratio}
\end{table}

\subsection{Emission and Absorption}
\label{sec:emission}
 The emission and absorption spectra of the scintillator cocktail were characterized in order to determine the Stokes' shift and understand the effect of self-absorption.

To determine the emission spectrum, LAB was first diluted in spectrophotometric-grade cyclohexane (2.5\,mL LAB per 1\,L cyclohexane) to minimize self-absorption. A PTI QuantaMaster\textsuperscript{TM} fluorescence spectrometer with an excitation wavelength of 250\,nm was then used to measure the emission spectrum for the diluted LAB, as is presented in Figure \ref{fig:emission}. As can be seen, LAB was measured to emit in the 275--350\,nm range.

\begin{figure}[htp]
  \centering
    \includegraphics[width=0.8\linewidth]{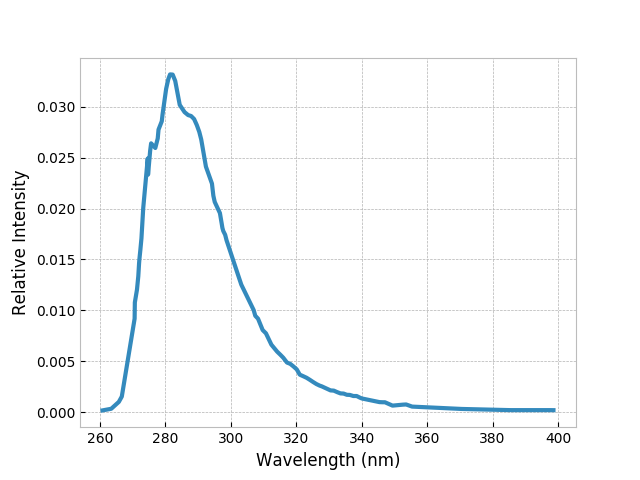}
  \caption{Emission spectrum of LAB, diluted to 2.5\,mL LAB per 1\,L cyclohexane to minimize self-absorption and excited using a 250\,nm fluorescence spectrometer.}
      \label{fig:emission}
\end{figure}

As mentioned in Section \ref{sec:snoplus} and more thoroughly discussed in Section \ref{sec:purification}, LAB was distilled in a purpose-built scintillator purification plant prior to deployment. The absorption spectrum of purified LAB was determined by placing the solution in a 1$\times$1\,cm quartz cuvette, which was measured between 190--1100\,nm using an Orion Aquamate 8000 spectrophotometer. The purified LAB was compared to the procured LAB after 0.2$\,\mu$m filtration but before distillation, as well as the best possible ``benchtop'' laboratory-purified LAB without using the purification plant. All measurements were performed simultaneously to control for instrumental systematics. The LAB purified by the plant had consistently superior optical clarity in the 330-400\,nm region of interest, as presented in Figure \ref{fig:lab_ug} (left).

The absorption and emission spectra of PPO have been extensively studied, and have previously been presented in compilations such as \cite{Berlman1971}. The SNO+ collaboration also measured the absorption and emission spectra of PPO, obtaining results consistent with these previous publications. The absorption spectrum of scintillator cocktail deployed within the SNO+ detector is shown in Figure \ref{fig:lab_ug} (right). As in Figure \ref{fig:lab_ug} (left), this spectrum is also compared to both undistilled and ``benchtop'' distilled LAB.  At wavelengths above which PPO is highly absorbing ($\sim$365\,nm), the superior optical clarity of the liquid scintillator is maintained after addition of the fluor.

\begin{figure}[htp]
    \includegraphics[width=0.48\textwidth]{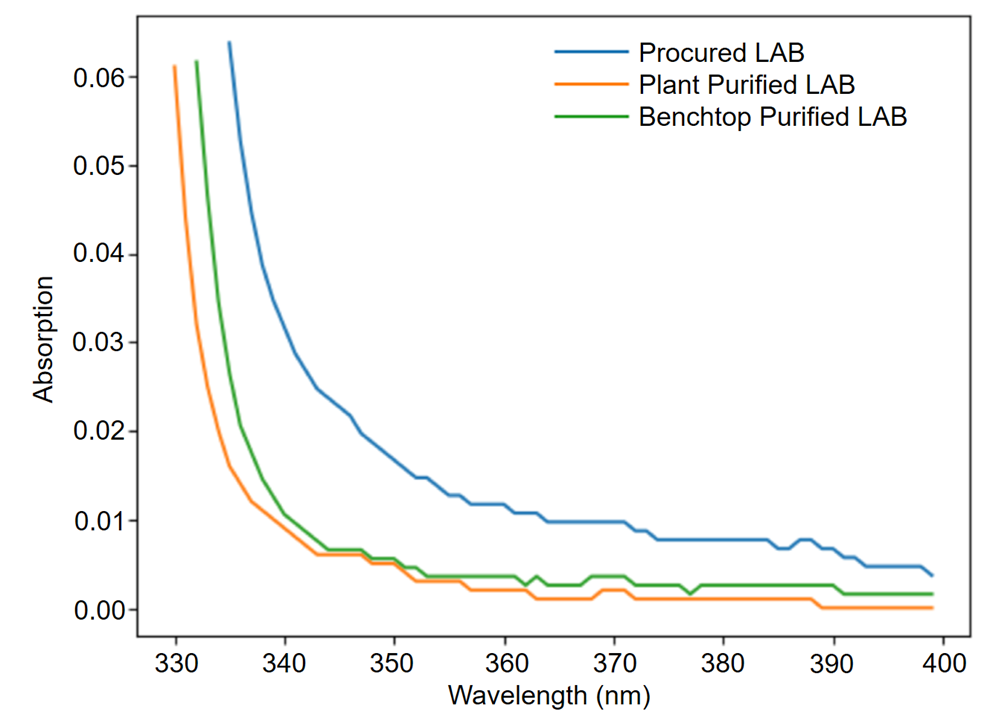}
	\includegraphics[width=0.48\textwidth]{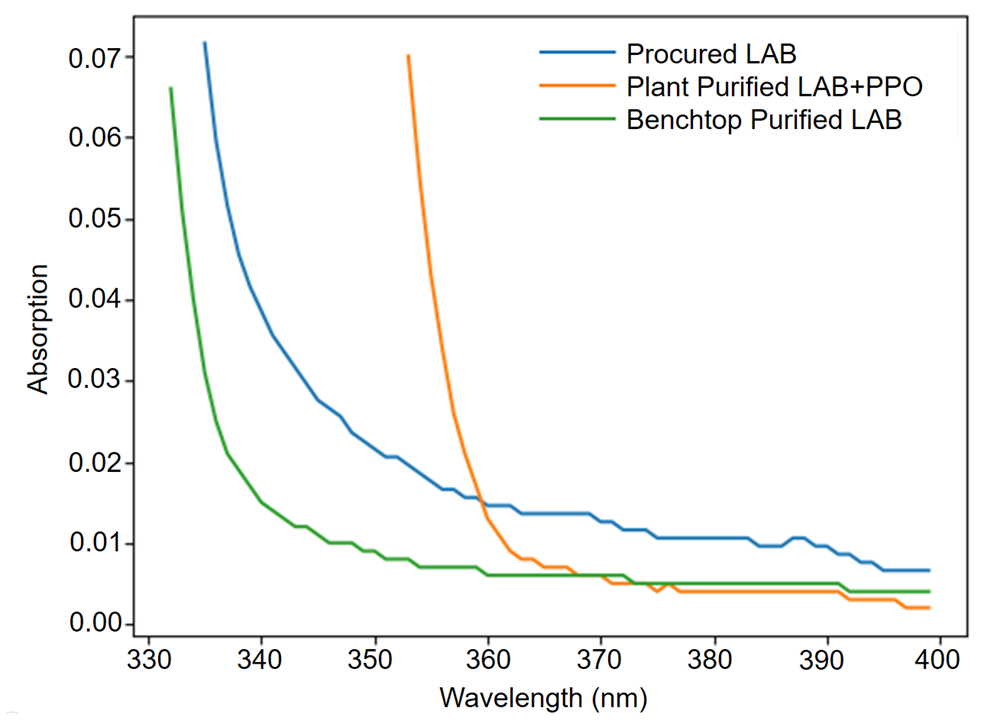}
    \caption{Absorption spectra of pure LAB and the scintillator cocktail (2\,g PPO per 1\,L LAB), produced by the SNO+ scintillator purification plant (orange), procured pure LAB after 0.2\,$\mu$m filtration but before distillation (blue) and best possible ``benchtop'' laboratory-purified pure LAB (green). Left: Pure LAB has superior optical clarity after distillation. Right: The scintillator cocktail has superior optical clarity after distillation at all wavelengths above the region where PPO is highly absorbing ($\sim$365\,nm). }%
    \label{fig:lab_ug}%
\end{figure}

\subsection{Transfer efficiencies}
\label{sec:transfer}

In order to determine the efficiency of the non-radiative transfer from LAB to PPO, the scintillator cocktail was optically excited in order to fit the decay time to the fluorescence response function for fluor `$y$', given a solvent `$x$' \cite{Birks1964},
\begin{equation}
n_{y}(t) = \frac{n_{0x}\frac{a_{xy}}{\tau _{txy}}}{\frac{1}{\tau_{0x}}-\frac{1}{(\tau_{0y})_{0}}}\left(e^{-\frac{t}{(\tau_{0x})_{0}}} - e^{-\frac{t}{\tau_{0x}}} \right).
\label{equ:birks}
\end{equation} 
Here, $n_{y}$ is the number of excited fluor molecules, $n_{0x}$ is the number of excited solvent molecules at $t=0$, $(\tau_{0y})_{0}$ is the molecular fluorescence time of the fluor (1.66\,ns for PPO \cite{Berlman1971}), $a_{xy}$ is the radiative transfer efficiency, and $\tau_{txy}$ is the non-radiative transfer decay time. $\tau_{0x}$ is the decay time for excitations of the solvent when in the cocktail, and $(\tau _{0x})_{0}$ is the decay constant of the pure solvent (measured using a time-based laser-excited fluorimeter to be 22.7\,ns for LAB).

Fitting Equation \ref{equ:birks} to extract $\tau_{0x}$, the non-radiative transfer efficiency can then be deduced using
\begin{equation}
f_{xy} = 1 - \frac{\tau_{0x}}{(\tau _{0x})_{0}}.
\end{equation}
The measurements of non-radiative transfer efficiency for LAB loaded with different concentrations of PPO are presented in Table \ref{table:nrt_timing}. 

\begin{table}[htp]
\centering
\begin{tabular}{||c| c||}
  
 \hline
 PPO Concentration & Non-Radiative Transfer Efficiency \\
(g\,PPO per 1\,L LAB) & \%\\
  \hline
4 &86.0 $\pm$ 0.8\\ 
2 &78.2 $\pm$ 1.5\\ 
1 &67.7 $\pm$ 2.3\\ 
0.5 &59.3 $\pm$ 3.2\\
0.25 &48.7 $\pm$ 5.0\\

 \hline
\end{tabular}
\caption{The non-radiative transfer efficiencies for various PPO concentrations in LAB.}
\label{table:nrt_timing}
\end{table}

\subsection{Re-emission}

Photons absorbed by the scintillator cocktail have a non-zero probability of being re-emitted. Measurements of the quantum yield (the emission probability of a fluorophore) typically involve comparing the fluorescence intensity of the sample against a known reference, such as quinine sulfate (C$_{20}$H$_{24}$O$_2$N$_2$). These measurements are prone to error, and literature values vary over a large range. This is due to a variety of parameters including oxygen quenching, wavelength dependencies, geometry, and dependence on the presence of the solvent matrix (in which numerous corrections must be included to extract accurate values) \cite{Buck2015}.

\paragraph{Absorption by PPO}
The quantum yield of PPO has been extensively measured with values ranging from 0.71--1.00. A recent paper reports a value of 0.842\,$\pm$\,0.042 for the quantum yield of PPO \cite{Buck2015}. When modelling re-emission time, light absorbed by PPO molecules is assumed to fluoresce with the intrinsic PPO exponential decay time constant of 1.6\,ns.

\paragraph{Absorption by LAB}

In cases where the photon is absorbed by an LAB molecule, the quantum yield is assumed to be $0.20\pm 0.02$, as measured by \cite{Buck2015}. However, direct re-emission by LAB is self-absorbing and does not contribute significantly to the total re-emission of the scintillator cocktail. As discussed in Section \ref{sec:transfer}, non-radiative energy transfer to a PPO molecule can occur with a probability of 0.782\,$\pm$\,0.015. Taking the product of the non-radiative transfer efficiency and the quantum yield of the PPO gives a calculated re-emission probability of 0.658\,$\pm$\,0.053 for a photon absorbed by an LAB molecule. 

When modelling re-emission, the dominant PPO emission spectrum is assumed. The re-emission time, following light absorption by LAB molecules and the subsequent re-emission by PPO after non-radiative transfer, was measured using UV fluorescence excitation to be 5.1\,$\pm$\,0.1\,ns.

\subsection{Timing Measurements}
\label{sec:timing}

The ability to discriminate between alpha particles ($\alpha$) and electrons ($\beta$) is an important capability for liquid scintillator detectors. Consequently, significant efforts were undertaken to parameterize the timing profile of the scintillator cocktail; these characteristics were also required for accurate modelling in Monte Carlo simulations.

When LAB is excited after interaction with an ionising particle, the excitation energy can be transferred to PPO through F{\"o}rster resonance energy transfer, a dipole-dipole non-radiative interaction. PPO then undergoes radiative decay, and de-excites from singlet or triplet states. Singlet states will decay directly to the ground state resulting in the prompt emission of scintillation light. De-excitation from the triplet to the singlet state is forbidden if intersystem crossing occurred, resulting in a delayed emission of scintillation light \cite{Birks1964}. The ratio between singlet and triplet states depend on the ionization density of the incident particle. Particles with high ionization densities, such as $\alpha$-particles, will produce a higher proportion of triplet states when compared to $\beta$-particles. Therefore, $\alpha/\beta$ discrimination can be achieved by using the liquid scintillator timing profile --- the intensity of scintillation light as a function of time.

Measurements have previously been made to determine the scintillation time profile of oxygenated and de-oxygenated LAB-based scintillator cocktails under $\alpha$ and $\beta$ excitation, with results published in \cite{OKeeffe2011}. The importance of de-oxygenation in the ability to perform $\alpha$/$\beta$ discrimination was demonstrated. However, subsequent measurements made by \cite{Lombardi2013, Li2011} were unable to replicate the very effective separation observed between $\alpha$ and $\beta$ events. As the setup and methodology were well understood and verified, it remains unclear why this particular result was seemingly better than other observations.

Due to this tension, two further complementary measurements were carried out by the SNO+ collaboration to characterize the timing profile of the scintillator cocktail. Both measurements used the single-photon sampling technique described in \cite{Bollinger1961} with slightly different methodologies. In each, a minimum of two PMTs were used; one to trigger scintillation events (``trigger PMT'') and another placed such that the average number of photons detected per trigger was less than one (``timing PMT''). The time difference between the trigger and timing PMTs yielded the emission time of the detected photon. The overall timing profile can then be built by accumulating these measurements over many events.

\paragraph{Method 1}

Measurements were taken by placing the liquid scintillator in a 3.5\,cm\,$\times$\,3.5\,cm\,$\times$\,3.5\,cm acrylic cube with a hollowed out, 2\,cm diameter, $\sim$11\,ml cylindrical sample cavity. The acrylic cube was optically coupled to a 1" trigger PMT, and a 30\,mm\,$\times$\,30\,mm timing PMT was placed at an offset of $\sim$30\,cm. $\beta$ events were produced using a $^{90}$Sr source while $\alpha$ events were produced using a $^{210}$Po source. Both sources were placed above the acrylic cube and separated from the cocktail by a $\sim$1\,mm air gap.

\paragraph{Method 2}
Measurements used three 2" cylindrical PMTs, and were taken by placing the liquid scintillator into an ultraviolet-transparent acrylic (UVT) sample holder. The sample holder was optically coupled to a trigger PMT. A secondary PMT was also coupled to the sample holder, which was deployed to understand the scintillator rise time. The timing PMT was masked and placed at a distance where it would only observe $<1\%$ of triggered events. $\beta$ events were produced using a $^{137}$Cs source placed beneath the sample holder, while $\alpha$ events were produced using a $^{244}$Cm source immersed within the scintillator cocktail.

\paragraph{Results}

The timing profiles of Method 1 were fit using a $\chi^{2}$-minimization approach while accounting for the dark rate and after-pulsing; the data was well described with a summation of four exponentials convolved with Gaussians. Conversely, the timing profiles of Method 2 were parameterized using a maximum likelihood fit while accounting for the dark current, single photon transit time of the PMTs, and trigger distribution; the data was also well described using summation of four exponentials. The fit parameters are presented in Table \ref{table:pennchicagotiming}, and the fits to the scintillation timing profiles are shown in Figure \ref{fig:timing_plots}. In order to examine the agreement between the two methods while controlling for experiment-specific effects, the scintillator cocktail timing parameters extracted from Method 2 were substituted into the Method 1 fit. Both fits were overlaid on the Method 1 data and presented in Figure \ref{fig:ab_comparison}. The corresponding $\alpha/\beta$ discrimination between the two new measurements are more comparable to the lower discrimination power published in \cite{Lombardi2013, Li2011} than those in \cite{OKeeffe2011}.

\begin{table}[htp]
\centering
\begin{tabular}{|| c | c | c | c | c ||} 
 \hline
 & \multicolumn{2}{ l | }{\hspace{30pt}Method 1} &\multicolumn{2}{ l || }{\hspace{40pt}Method 2}\\
  \hline

 Fit & $\alpha$ & $\beta^{-}$ & $\alpha$ & $\beta^{-}$\\
  \hline
$\tau_{1}$\,(ns) & 4.66$\pm$0.17 & 5.1$\pm$0.2& 4.79$\pm$0.18 & 4.88$\pm$0.08 \\
$\tau_{2}$\,(ns) & 14.2$\pm$2.0 & 17.6$\pm$2.0 & 18.4$\pm$0.9 & 15.4$\pm$0.7\\
$\tau_{3}$\,(ns) & 64.3$\pm$6.1 & 45.3$\pm$5.1 & 92.0$\pm$5.0 & 66.0$\pm$4.0\\
$\tau_{4}$\,(ns) & 578$\pm$313 & 498$\pm$50 & 900$\pm$110 & 400$\pm$40\\
A$_{1}$ & 0.44$\pm$0.02 & 0.66$\pm$0.02 & 0.427$\pm$0.01 & 0.665$\pm$0.008\\
A$_{2}$ & 0.31$\pm$0.01 & 0.20$\pm$0.01 & 0.313$\pm$0.01 & 0.218$\pm$0.009\\
A$_{3}$ & 0.16$\pm$0.01 & 0.08$\pm$0.01 & 0.157$\pm$0.06 & 0.083$\pm$0.004\\
A$_{4}$ & 0.09$\pm$0.02 & 0.06$\pm$0.02 & 0.102$\pm$0.003 & 0.035$\pm$0.002\\

 \hline
\end{tabular}
\caption{The fit parameters for the $\alpha$ and $\beta$ timing profiles of the scintillator cocktail. $\tau_{i}$ are the decay constants and $A_{i}$ are the scaling factors for each exponential.}
\label{table:pennchicagotiming}
\end{table}

\begin{figure}[htp]
  \centering
    \hspace{-1.35cm}
    \includegraphics[width=0.75\linewidth]{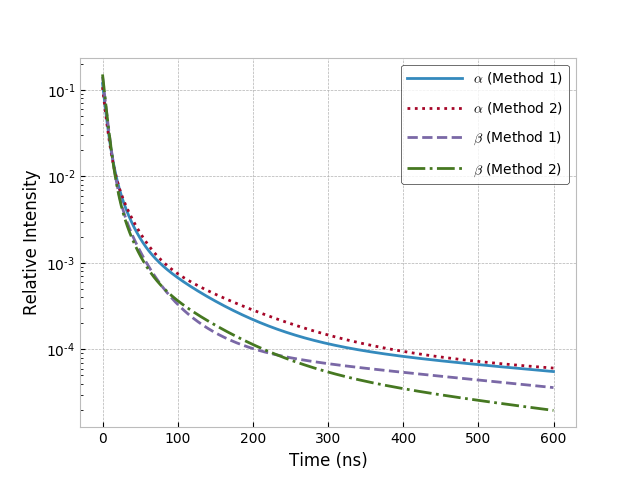}
  \caption{The fits for $\alpha$ and $\beta$ timing profiles for both Method 1 and Method 2. Each fit used a summation of four exponentials convolved with Gaussians, with fit values displayed in Table \ref{table:pennchicagotiming}.}
      \label{fig:timing_plots}
\end{figure}

\begin{figure}[htp]
    \includegraphics[width=0.50\textwidth]{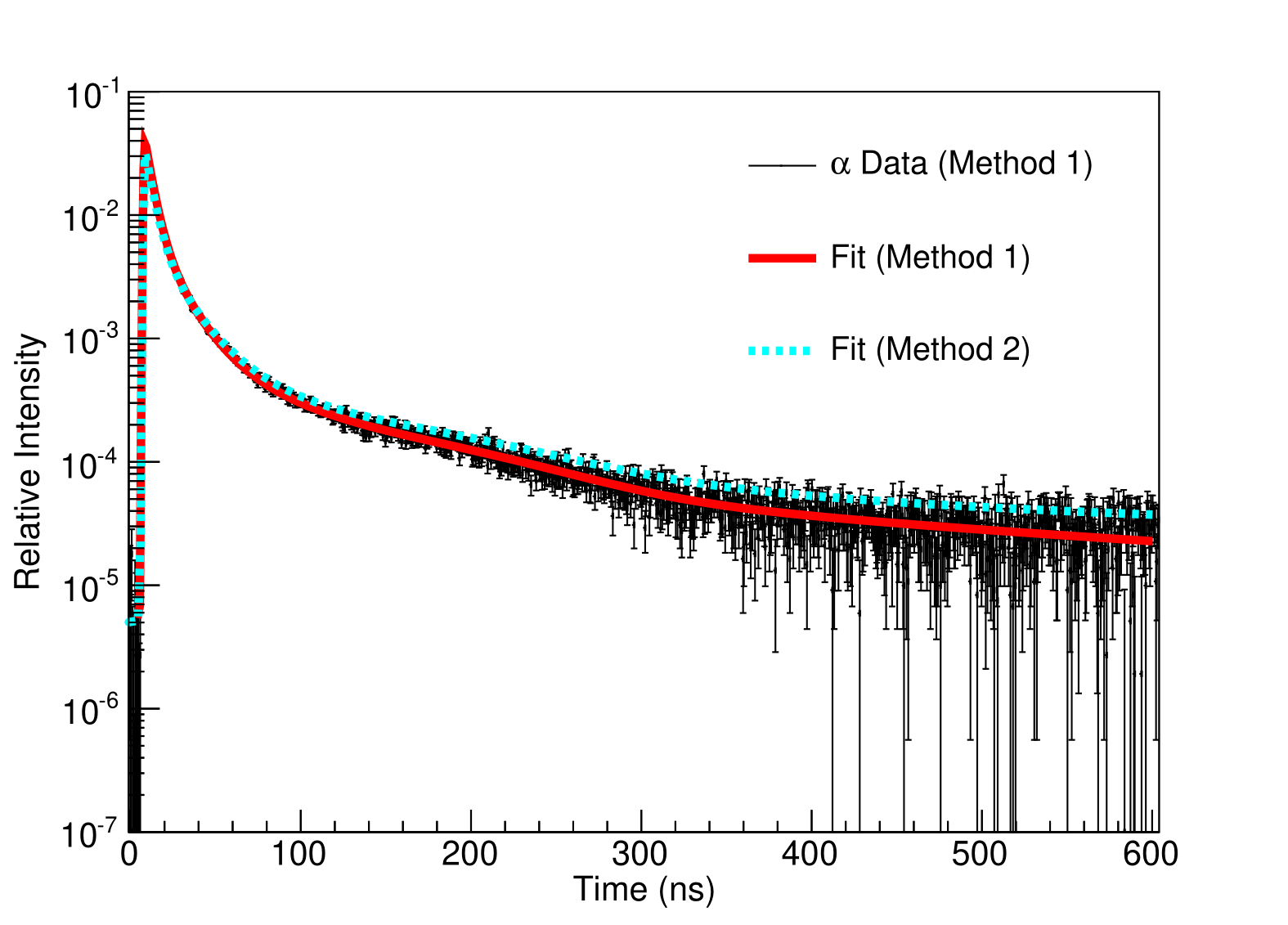}
	\includegraphics[width=0.50\textwidth]{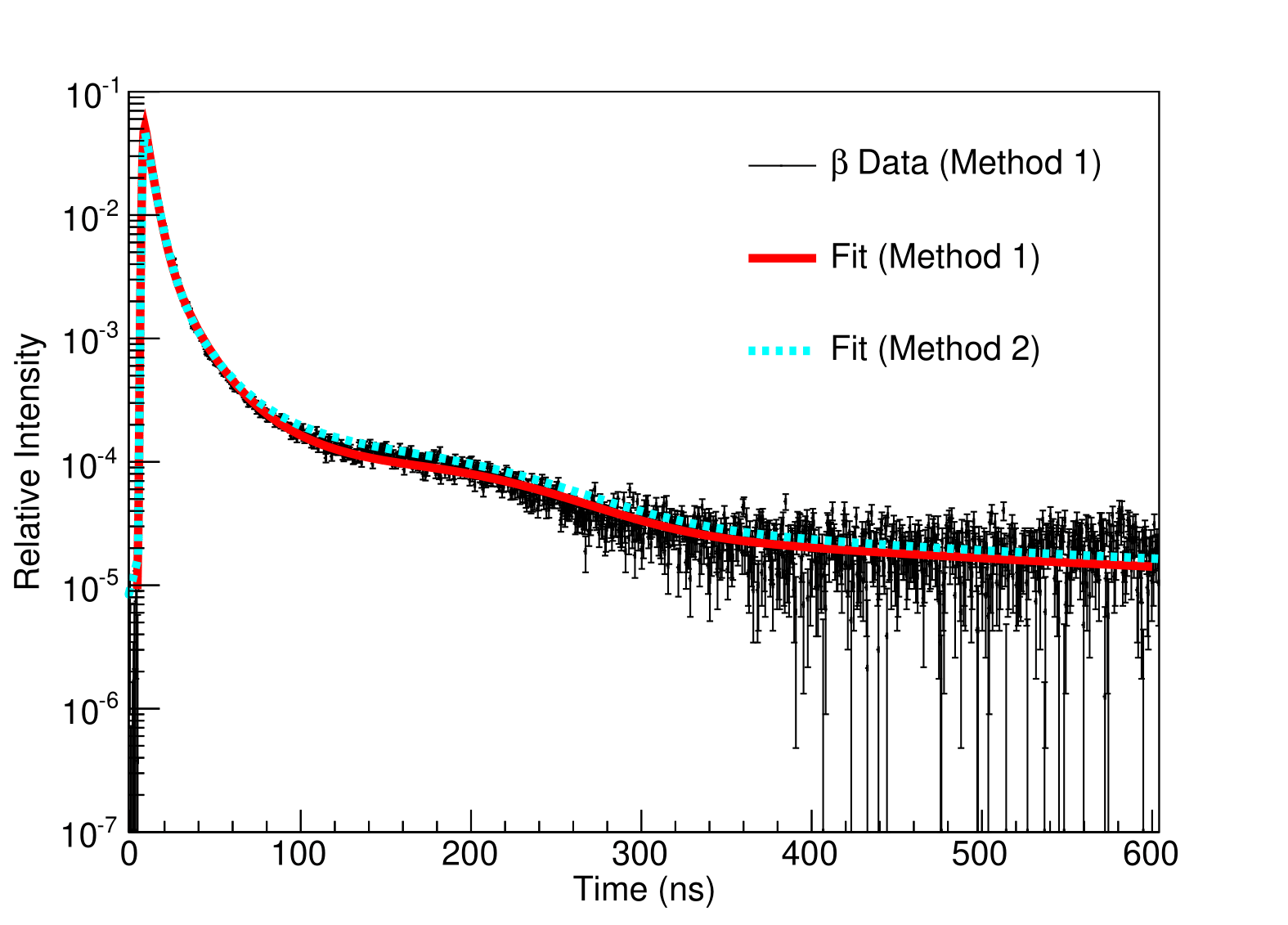}
    \caption{Data taken using Method 1, fit using a $\chi^{2}$-minimization approach while accounting for detector effects such as the dark rate and after-pulsing. The scintillator cocktail timing parameters from Method 2 were substituted into the Method 1 fit and overlaid on the Method 1 data. Left: $\alpha$ timing profile. Right: $\beta$ timing profile.}%
    \label{fig:ab_comparison}%
\end{figure}

\subsection{Refractive Index}

The relationship between refractive index and wavelength in LAB+PPO was first published in \cite{Tseung2011a} using 3.0\,g PPO per 1\,L LAB. A Woollam M2000 ellipsometer was used to scan wavelengths of this higher concentration scintillator cocktail between 210\,-\,1000\,nm at 5 incident angles (55$^{\circ}$, 60$^{\circ}$, 65$^{\circ}$, 70$^{\circ}$ and 75$^{\circ}$). The average of all 5 incident angles was used to determine the refractive index at each wavelength. As shown in Figure \ref{fig:ref_i}, these results were compared to other measurements made by SNO+ and RENO \cite{Yeo2010}. More details about this measurement can be found in \cite{Tseung2011a}.  Based on other measurements that were performed, these results are also applicable at the fluor concentration of the scintillator cocktail (2\,g PPO per 1\,L LAB).

\begin{figure}[htp]
  \centering
    \hspace{-1.35cm}
    \includegraphics[width=0.65\linewidth]{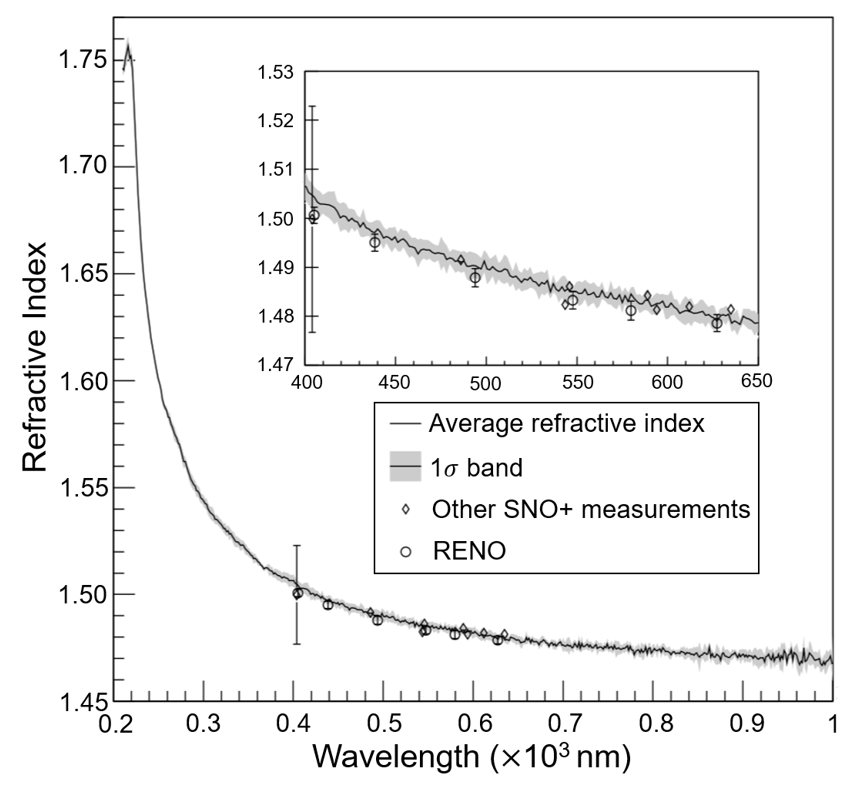}
  \caption{The wavelength dependent refractive index of the scintillator cocktail with a higher PPO loading of 3.0\,g/L. The shaded region represents 1$\sigma$ deviations. These results are compared to previous refractometry measurements by SNO+ and RENO \cite{Yeo2010}. Figure taken from \cite{Tseung2011a}.}
      \label{fig:ref_i}
\end{figure}

\subsection{Density}

The density of the scintillator cocktail depends on temperature. To characterize this, samples of both pure LAB and the scintillator cocktail were either placed in glass syringes and cooled in a refrigeration unit, or placed in a glass beaker and heated on a hot plate. The samples were then measured using a PAAR DMA 35 densitometer. As shown in Figure \ref{fig:density}, the density of pure LAB and the scintillator cocktail was assessed between 10--26$^{\circ}$C. These measurements are comparable with those published in \cite{Zhou2015b}.

\begin{figure}[htp]
  \centering
    \includegraphics[width=0.75\linewidth]{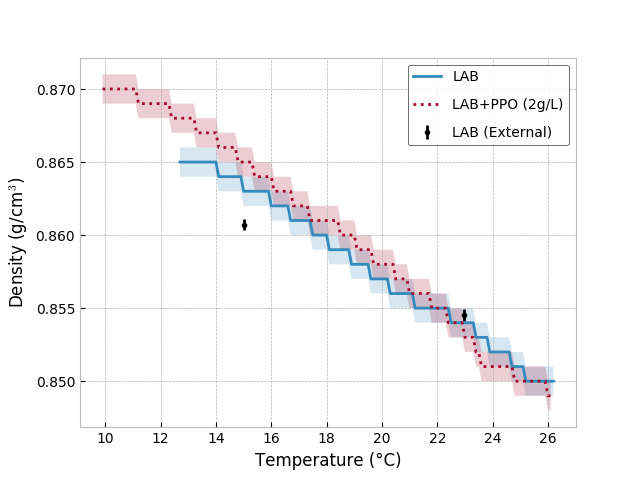}
  \caption{Comparison of density variation of the scintillator cocktail between 10--26$^{\circ}$C. The shaded regions represent instrumental uncertainties from the densitometer. The black points were measurements published in \cite{Zhou2015b}.}
      \label{fig:density}
\end{figure}
	\section{Scintillator Cocktail Deployment}
\label{sec:purification}
The SNO+ scintillator production and purification methodology was previously discussed in \cite{Ford2015}. This section will provide an overview of the production, purification and quality assurance processes used during the deployment of the scintillator cocktail. The technical details of these deployment processes will be further elaborated in a forthcoming paper following \emph{in-situ} measurements of the deployed scintillator cocktail.

In order to meet the strict cleanliness and purity requirements of the experiment \cite{detectorpaper}, P\,500-Q\,LAB was purchased from CEPSA Qu\'imica B\'ecancour Inc. Upon receiving shipments, the ultraviolet-visible (UV-Vis) and Fourier-transform infrared (FTIR) spectra, density, and turbidity were measured. If the delivered LAB had sufficient quality as deemed by these measurements, it was stored in a stainless steel holding tank located in the SNOLAB surface facility. This tank was previously passivated with citric acid and cleaned, and the LAB stored within was kept under a blanket of high purity nitrogen gas. The LAB was then loaded onto stainless steel railcars, each cleaned in the same manner as the surface storage container. These railcars were subsequently transported underground, where the LAB was unloaded into two storage tanks, each capable of holding 50 tonnes of LAB. These storage tanks have a polyproylene liner, and were cleaned in accordance with MIL-STD-1247C level 50 using both a 1\% Alconox detergent solution and ultrapure water.\footnote {MIL-STD-1247C is the United States Military Standard on Product Cleanliness Levels and Contamination Control Program. \href{http://everyspec.com/MIL-STD/MIL-STD-1100-1299/MIL_STD_1246C_131/}{www.everyspec.com/MIL-STD/MIL-STD-1100-1299/MIL\_STD\_1246C\_131/}.}

Once underground, the LAB was purified using the SNO+ scintillator purification plant. The first step of this purification was a continuous multi-stage distillation. This system heats the LAB to 220$^{\circ}$C at an absolute pressure of 40\,Torr.  The distillation column was refluxed such that a distillate flow rate of up to 750\,kg/hr was achieved. This process removed lower volatility impurities including heavy metals (Bi, K, Pb, Po, Ra and Th) and oxidised organics (carboxyl groups and 1,4-benzoquinone). The distilled LAB then underwent N$_{2}$ stripping at 100$^{\circ}$C and an absolute pressure of 150\,Torr. This process removed dissolved gases (Ar, Kr, O$_{2}$ and Rn) and volatile liquids (such as residual water). The capability to perform steam stripping on LAB was developed, and may be used to polish the deployed scintillator cocktail in the future.

The distilled and N$_{2}$-stripped LAB was systematically monitored throughout all purification processes. Improvements to the solvent were assessed using UV-Vis spectroscopy, densitometry, and nephelometry. These diagnostic tools ensured that the purification plant performed as expected, with the assumption that any diagnostic anomalies implied a failure in the process systems and consequently a potential ingress of contamination. Tested every hour, only LAB that passed this strict quality assurance regime were utilised in further stages. An example of the UV-Vis spectra performed during this verification process is shown in Figure \ref{fig:lab_ug}.

The PPO used in the deployed scintillator cocktail was purchased as ``neutrino grade'' from PerkinElmer, delivered to SNOLAB in plastic drums, and shipped directly to the underground laboratory. The PPO was mixed with purified LAB to a target concentration of $\sim$120\,g PPO per 1\,L LAB to produce a high-concentration ``master solution''. This master solution was then sparged three times with high purity N$_{2}$ gas using a vacuum pump/purge system to remove dissolved gases such as O$_{2}$ and Rn. Three solvent-solvent water extractions were then performed on the master solution, in which the raffinate was mixed with ultrapure water and allowed to re-separate. This removed U, Th, Ra, K, and Pb, as well as other charged particulates and ionic impurities. The master solution was subsequently filtered and distilled in a single-stage kettle at 230$^{\circ}$C and an absolute pressure of 20\,Torr. At this temperature and pressure, the LAB flashed through the system while the PPO was distilled, and the two components were re-mixed in a condenser.

The distilled master solution was mixed in-line with purified LAB to attain the desired scintillator cocktail concentration of 2\,g PPO per 1\,L of LAB. This was then passed through 0.05\,$\mu$m filters before being systematically tested every hour using UV-Vis spectroscopy, densitometry and nephelometry. Further tests were also regularly performed; these included light yield, FTIR, and neutron activation analysis. Once the purity of the scintillator cocktail was assured through a multi-stage approval process, the cocktail was cooled to a target temperature of 12\,$^{\circ}$C and sent to the SNO+ AV.

Following deployment of the scintillator cocktail within the SNO+ detector, the purification plant is designed to maintain the optical clarity and radiopurity of the scintillator through solvent-solvent extraction with ultrapure water, steam stripping, and the use of metal scavengers. These processes are designed to operate at a flow rate of $\sim$130\,litres per minute, allowing for a recirculation of the entire 780\,tonne detector volume in $\sim$100\,hours, comparable to the half-life of $^{222}$Rn.
	\graphicspath{{./}{figures/}}
\section{Summary}

Due to the incompatibility between existing widely-used liquid scintillators and the acrylic used to construct the SNO+ detector, the SNO+ collaboration developed a new liquid scintillator. The resulting scintillator cocktail of 2\,g PPO per 1\,L LAB is not only compatible with acrylic, but exhibits a competitive light yield to existing liquid scintillators while maintaining other advantages including longer attenuation lengths, superior safety characteristics, chemical simplicity, ease of handling, and logistical availability to the SNO+ site.

In preparation for the SNO+ scintillator phase physics program, properties of the scintillator cocktail were extensively characterized to ensure that the detector is accurately modelled in Monte Carlo simulations. The LAB-based liquid scintillator developed by the SNO+ collaboration is now used in multiple large-scale experiments, and has been deployed in the SNO+ experiment.
	\endgroup	
	
 	\section*{Acknowledgments}
\label{sec:acknowledgements}

Capital construction funds for the SNO+ experiment were provided by the Canada Foundation for Innovation (CFI) and matching partners.
This research was supported by:
{\bf Canada}: Natural Sciences and Engineering Research Council, the Canadian Institute for Advanced Research (CIFAR), Queen’s University at Kingston, Ontario Ministry of Research, Innovation and Science, Alberta Science and Research Investments Program, National Research Council, Federal Economic Development Initiative for Northern Ontario (FedNor), Northern Ontario Heritage Fund Corporation, Ontario Early Researcher Awards, Arthur B. McDonald Canadian Astroparticle Physics Research Institute;
{\bf Germany}: the Deutsche Forschungsgemeinschaft (DFG);
{\bf Mexico}: DGAPA-UNAM and Consejo Nacional de Ciencia y Tecnolog\'{i}a;
{\bf Portugal}: Funda\c{c}\~{a}o para a Ci\^{e}ncia e a Tecnologia (FCT-Portugal);
{\bf UK}: Science and Technology Facilities Council (STFC), the European Union’s Seventh Framework Programme under the European Research Council (ERC) grant agreement, the Marie Curie grant agreement;
{\bf US}: Department of Energy Office of Nuclear Physics, National Science Foundation, the University of California, Berkeley, Department of Energy National Nuclear Security Administration through the Nuclear Science and Security Consortium.

We would like to thank SNOLAB and its staff for support through underground space, logistical, and technical services.
SNOLAB operations are supported by the Canada Foundation for Innovation and the Province of Ontario, with underground access provided by Vale Canada Limited at the Creighton mine site.



    
    \bibliographystyle{bib/JHEP}
    \bibliography{bib/references}

\end{document}